# Title : Exergy – a useful concept for ecology and sustainability

Author : Göran Wall and Dilip G. Banhatti

Affiliation and postal address for communication:
Dr. Dilip G. Banhatti            |     until about June-end 2012
School of Physics                |     Dr. Dilip G. Banhatti
Madurai Kamaraj University       |     Hittorfstrasse 49a
Madurai 625021                   |     D-48149 Muenster
India                            |     Germany

E-mail address : dilip.g.banhatti@gmail.com

[Telephonically not accessible until about June-end 2012.]

# Exergy – a useful concept for ecology and sustainability

Göran Wall[1] and Dilip G. Banhatti[2],
(1) Solhemsgatan 46, SE-431 44 Mölndal, Sweden
(2) School of Physics, Madurai Kamaraj University, Madurai 625021, India

**Abstract:** We present the relatively less known thermodynamic concept of exergy in the context of ecology and sustainability. To this end, we first very briefly outline thermodynamics as it arose historically via engineering studies. This enables us to define exergy as available energy. An example of applying the concept of exergy to a simple human process is next described. Then we present an exergy analysis of Earth as a flow system, also concurrently describing other necessary concepts. Finally, we briefly comment on the applicability of exergy analysis to ecology and sustainability.

**Keywords:** exergy – energy – work – thermodynamics – heat transfer – flow processes

**THERMODYNAMICS – BRIEF HISTORY**

In early 19$^{th}$ century, design and construction of steam engine led to industrial revolution. Efforts to understand working of steam engine quantitatively laid foundations of the science of thermodynamics. Thermodynamics defines, prescribes how to measure, and relates to each other heat Q, temperature T, internal energy U, pressure p, volume V, density ρ, work W, entropy S and so on, especially as applied to the working substance of a heat engine like steam engine. Of these thermodynamic quantities, Q, U, V and S are called extensive properties since they are proportional to the amount of substance, while T, p and ρ are intensive, as they depend only on the state of the substance, independent of the amount. Work W is a measure of directed motion – of the steam engine piston, for example. By mid-19$^{th}$ century, widespread application and theoretical analysis led to absolute scale of temperature T, and the absolute zero (T = 0 K) on this scale, the coldest possible. By considering changes in the state of a system, and applying the already well developed classical mechanics, first law of thermodynamics extended conservation of mechanical energy to include as a form of energy the heat dQ transferred to the working substance of a heat engine at temperature T.

Thus, for a heat engine, which is defined as a flow process designed to convert heat to work,
$$dQ = dU + dW = dU + pdV.$$

Heat dQ transferred to the substance increases its internal energy by dU and moves the piston through volume dV at pressure p, thus doing mechanical work dW = pdV. This balance of dQ with dU + dW is a statement of conservation of energy, also called first law of thermodynamics. Work dW must maximize for the greatest motive power, and the maximum possible work is obtained by minimum generation of entropy dS = dQ/T. Thus,
$$dQ = TdS = dU + dW.$$

In addition, the entropy of the ambient medium at temperature $T_0$ surrounding the heat engine also changes, both together increasing in the real irreversible process that takes place. The minimal increase of entropy of system + surroundings leads to maximum work and hence maximum efficiency η of energy conversion from heat to work. This is given by

$$\eta = 1 - T_0/T,$$

called Carnot efficiency, independent of working substance. Existence of this limiting efficiency is a statement of second law of thermodynamics, which can also be expressed as

$$dS \text{ (system + surroundings)} > 0.$$

Towards end of 19th century, empirical thermochemistry was theoretically systematized and incorporated into thermodynamics, thus including chemical energy in addition to mechanical. This needed the concept of chemical potential µ, an intensive property, in combination with number of moles M of the chemical substance being considered. M is an extensive property. Energy conservation had already been extended from classical mechanics to other areas of physics like electromagnetism. Thermodynamics naturally used electric and magnetic potentials and energies when applied to such systems. Efforts to derive the macroscopic principles of thermodynamics from atomic structure of matter gave rise to statistical physics. As physics developed further into relativistic and quantum realms, these methods and results broadened the scope of thermodynamics further. For a more detailed history relevant for exergy, see articles listed at [http://exergy.se], available as pdf files.

**EXERGY**

Exergy is formed from Greek ex + ergon, meaning "from work". Some synonyms (from Wikipedia) are: availability, available energy, exergic energy, essergy, utilizable energy, available useful work, maximum (or minimum) work or work content, reversible work, and ideal work. ***Exergy is that part of energy which is convertible into all other forms of energy.*** It represents the potential of a system to deliver work in a given environment. The exergy E of a system of volume V having internal energy U and entropy S, and composed of many substances i (i = 1, 2, …), each amounting to $M_i$ moles (and having chemical potential $\mu_{0i}$ in the surroundings), is defined as

$$E = U + p_0 V - T_0 S - \sum_i \mu_{0i} M_i = H - T_0 S - \sum_i \mu_{0i} M_i$$

relative to surroundings with pressure $p_0$ and temperature $T_0$. H is a derived thermodynamic property called enthalpy. For flow in an open steady state system, it includes, in addition, the kinetic energy (per unit mass) $(1/2)\rho v^2$ of the flow of fluid of density ρ and speed v. For the system, the internal energy U is given by

$$U = TS - pV + \sum_i \mu_i M_i,$$

where T, p and $\mu_i$ refer to properties of the system. Using this expression for U, exergy E can also be written as

$$E = S(T - T_0) - V(p - p_0) + \sum_i (\mu_i - \mu_{0i})M_i.$$

This clearly shows that exergy is measured relative to a reference environment. For more details and relation of exergy to other thermodynamic quantities like Gibbs free energy, Helmholtz free energy, reference may be made to G. Wall (1977 Exergy – a useful concept within resource accounting, available as a pdf file from [http://exergy.se]). The same paper applies exergy to evaluate quality of substances like ores and others relative to Earth's average environment. This application of exergy has been developed further over the decades. There is as yet no consensus on its unambiguous use for this purpose – see references [1]-[3] at the end for a discussion of this aspect.

**HUMAN PROCESSES**

The concept of exergy has been successfully used in engineering for assessing and improving various types of plants and processes – see references [4]-[16]. These include energy generation, manufacture of consumer goods, acclimatization units for housing, equipment utilized for agriculture, metallurgical processes to extract metals from ores and so on. As an example, a thermal power plant is displayed and discussed briefly below.

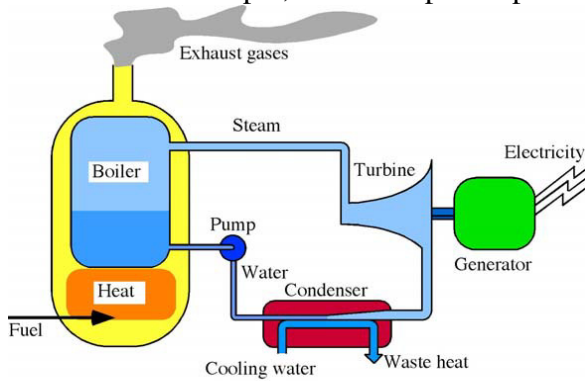
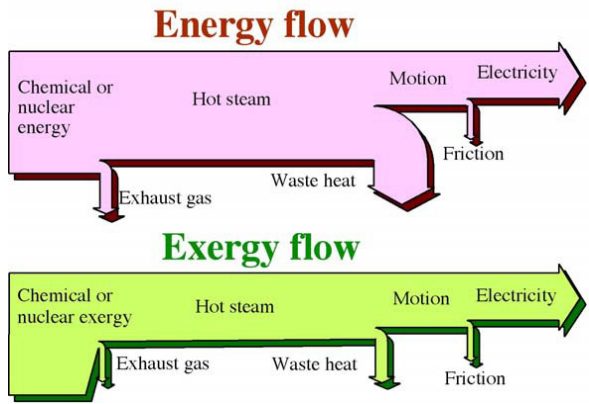

-------------------------------------------------------
Energy and exergy flows through a condensing thermal power plant.
**(From G Wall & M Gong 2001 Exergy Int. J. 1(3) 128-45 On exergy & sustainable development – Part 1 – Conditions & Concepts – p. 137 – Cf [http://exergy.se].)**
-------------------------------------------------------

This figure schematically shows a generic thermal power plant and compares and contrasts the energy and exergy flows through it. The widths of the thick arrows are proportional to the amounts of flow. The energy and exergy losses are clearly seen and can be compared for the same input and output. Arrows which turn downward indicate disposal to the environment. There is no narrowing of arrows in energy flow, while exergy arrow becomes narrow due to exergy loss to irreversibility. Thus boiler has largest irreversibility. So improving it will enhance efficiency most.

**NATURAL PROCESSES**

The grandest natural process consists of (our current understanding of) birth and evolution of universe in big bang theory on the timescale of 14 Gyr. Formation of Earth as a planet of our solar system occurred about 5 Gyr ago. Geological orogenic (*i.e.*, mountain building) cycles occur every 300 to 500 Myr in Earth's evolution since its consolidation. There are also natural processes of a few Kyr to decades (*e.g.*, nitrogen cycle and other climatic and ecological cycles), down to 1 yr (hydrologic cycle). In the biosphere, there are processes over millennia, centuries, decades, years, months, weeks, days and shorter, down to milliseconds, and perhaps even shorter cellular (sub)processes in biological cells.

Exergy to drive all natural processes on Earth has its origin in sunlight. This is illustrated in a series of schematic diagrams and figures below, each with brief comments.

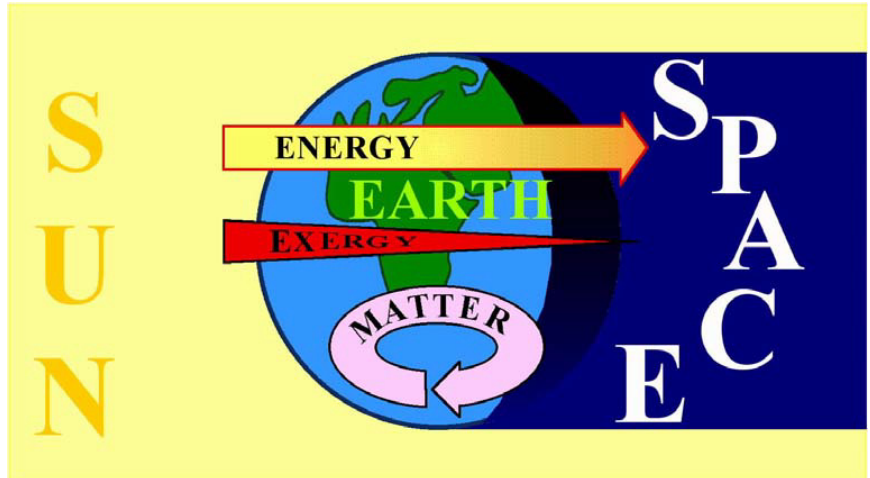

------------------------
The Sun-Earth-space system.
**(From Göran Wall preprint 2011 Exergy and Sustainable Development p. 7 – Cf [http://exergy.se].)**
------------------------

Solar energy flows to Earth and out again without loss, while exergy is partially lost and partially stored via mainly the process of photosynthesis, while matter undergoes upheaval on many space and time scales from geologic to much smaller. Many bio- and geo-processes are involved.

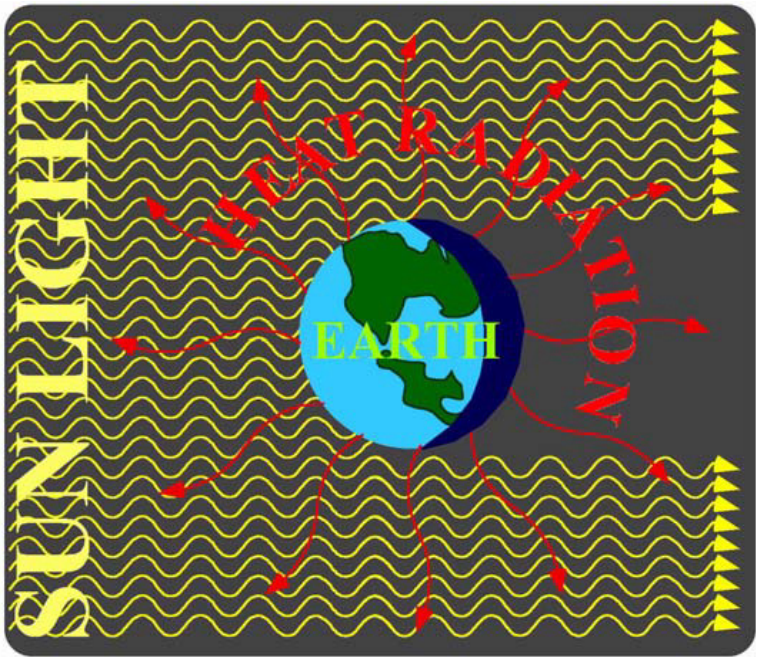

------------------------------------
Short-wave sunlight inwards and long-wave heat radiation outwards.
**(From Göran Wall preprint 2011 Exergy and Sustainable Development p. 8 – cf [http://exergy.se].)**
------------------------------------

Although all energy incident on Earth from Sun is radiated away, quality of incoming energy, measured by its exergy, is much higher than what is radiated away. Thus, incident energy equals radiated energy, but exergy content of incident energy is larger than the exergy content of radiated energy. Many symbiotic interconnected systems and processes are involved in the details. Over the last few decades, these have been described in detail in an overarching concept named Gaia after ancient Greek goddess of Earth. For details, the reader is referred to James Lovelock's book (1988, 1995 Norton) *The Ages of Gaia: a biography of our living Earth*. The relevance of such whole Earth studies for mitigating rapidly accelerating climate change cannot be overemphasized. We also refer readers to presentations at another Humboldt Kolleg held in Salem (TamilNadu, India) in September 2011 on a related theme abbreviated Earth-Future 2011.

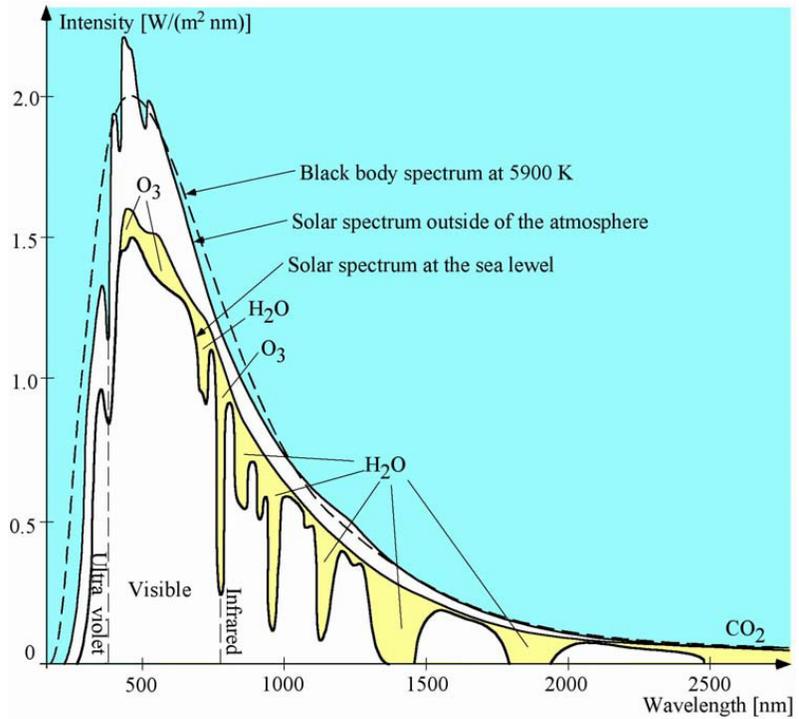

The intensity of sunlight at the surface of the Earth with respect to wavelength.
**(From Göran Wall preprint 2011 Exergy and Sustainable Development p. 8 – cf [http://exergy.se].)**

This figure shows spectrum of sunlight, *i.e.*, intensity as a function of wavelength. The dotted line shows ideal sunlight, the curve closest to it shows sunlight incident on top of Earth's atmosphere, while other curves show absorption bands due to various constituents of Earth's atmosphere, as labelled.

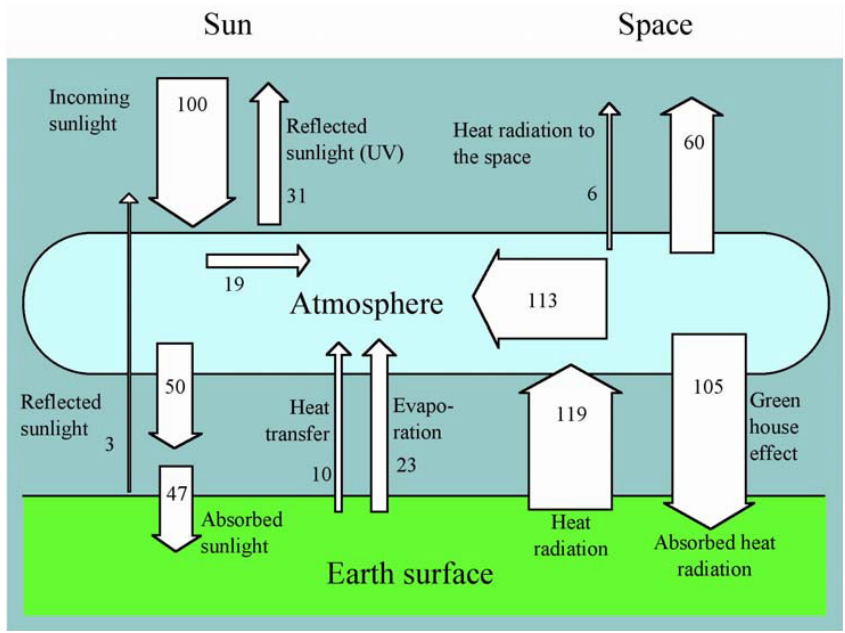

The energy flow between the Sun, the atmosphere, the surface of the Earth and space. **Numbers are in percent of incoming sunlight.**
**(From Göran Wall preprint 2011 Exergy and Sustainable Development p. 9 – cf [http://exergy.se].)**

A representation of energy audit of sunlight incident on Earth and partially used and stored, before some part is radiated away. This has remained more or less steady over aeons of time since Earth acquired an oxygen dominated atmosphere some 3 Gyr ago. Evidence for this near constancy comes from paleogeochronological indicators of various kinds as gauged by radioactive dating methods.

---

The global exergy flows on the Earth, where 1x is equal to $1.2 \times 10^{13}$ W.
(From Göran Wall preprint 2011 Exergy and Sustainable Development p. 10 – cf [http://exergy.se].)

---

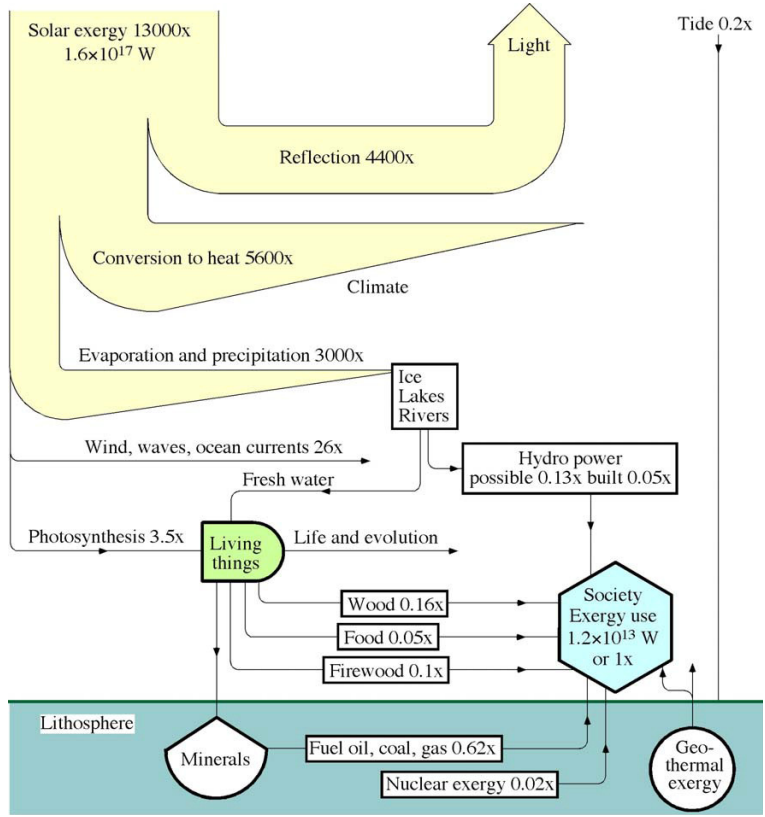

Another representation of exergy audit of Earth. There is no unanimity if materials can be assigned exergy, nor agreement on what should be taken as the reference state for exergy calculations. Even then such evaluations can be helpful in thinking globally about Earth matters. At the present juncture in human history, such thinking and proactive measures are very essential for mitigating accelerated climate change.

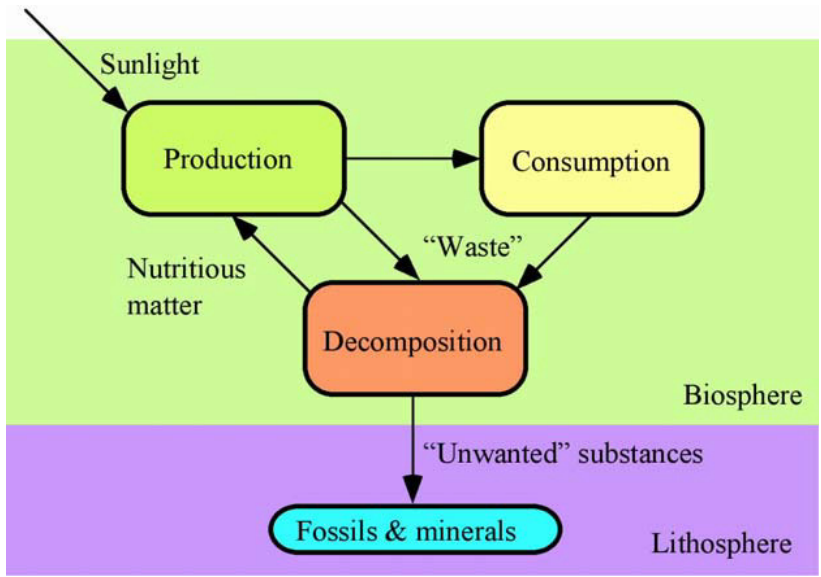

---

The circulation of matter in nature is powered by sunlight.
(From Göran Wall preprint 2011 Exergy and Sustainable Development p. 11 – cf [http://exergy.se].)

---

A schematic diagram for movement of matter on Earth. Bio- and geo-processes on many different time and space or size scales are involved in the details behind this sketch.

The next few sketches or diagrams indicate, by classifying resources of Earth, how human use of these has been unsustainable after industrial revolution ushered in by steam engine, and accelerated very rapidly of late.

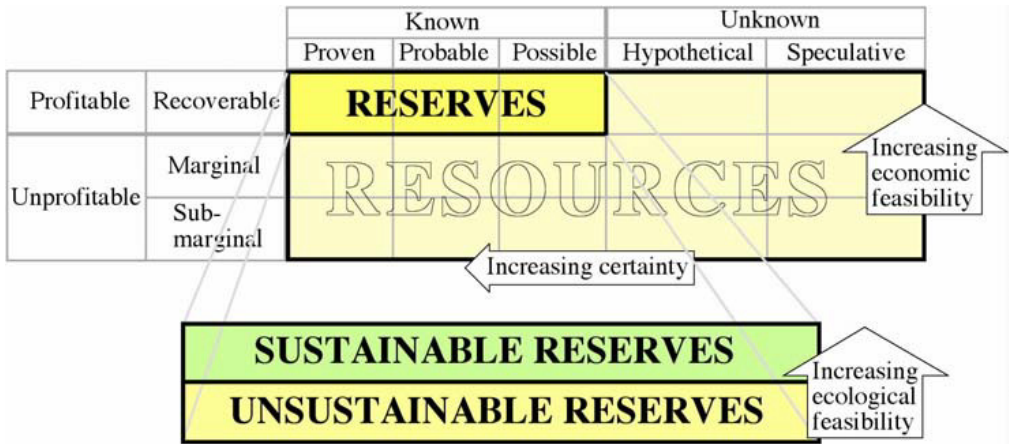

---
Definition of resources and reserves
**(From Göran Wall preprint 2011 Exergy and Sustainable Development p. 18 – cf [http://exergy.se].)**

---

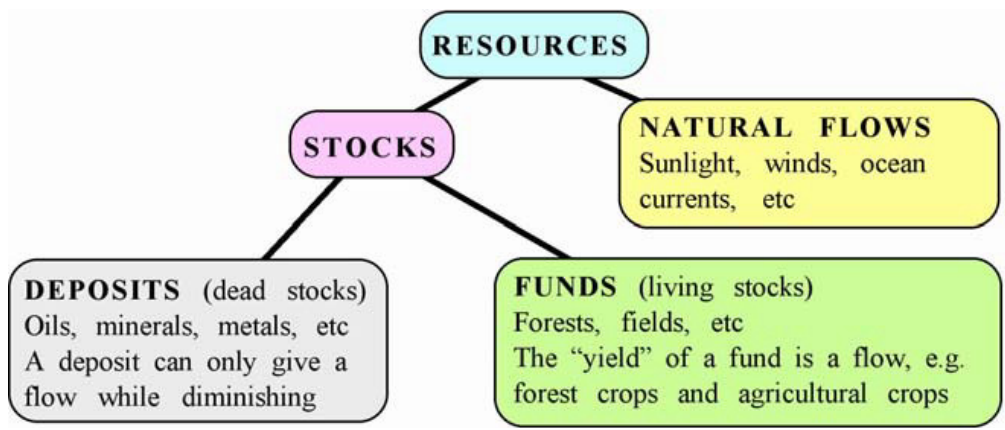

---
Classification of resources
**(From Göran Wall preprint 2011 Exergy and Sustainable Development p. 18 – cf [http://exergy.se].)**

---

Society takes deposits from nature and returns wastes
**(From Göran Wall preprint 2011 Exergy and Sustainable Development p. 19 – cf [http://exergy.se].)**

---

Current unsustainable pattern of human societal use of vast but limited natural resources. The deposits take aeons to build up, but are used up within centuries at most, at current rate of use.

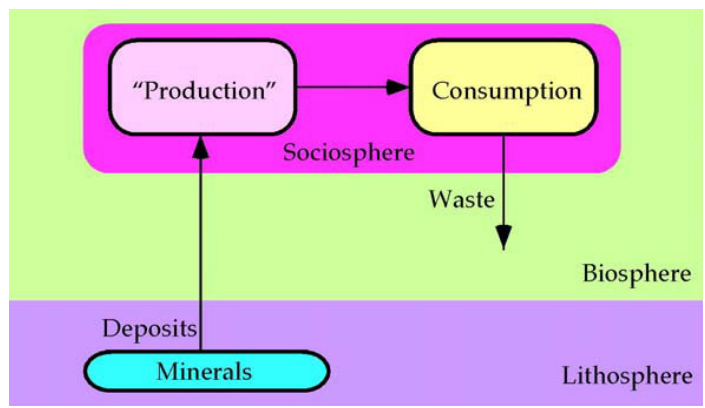

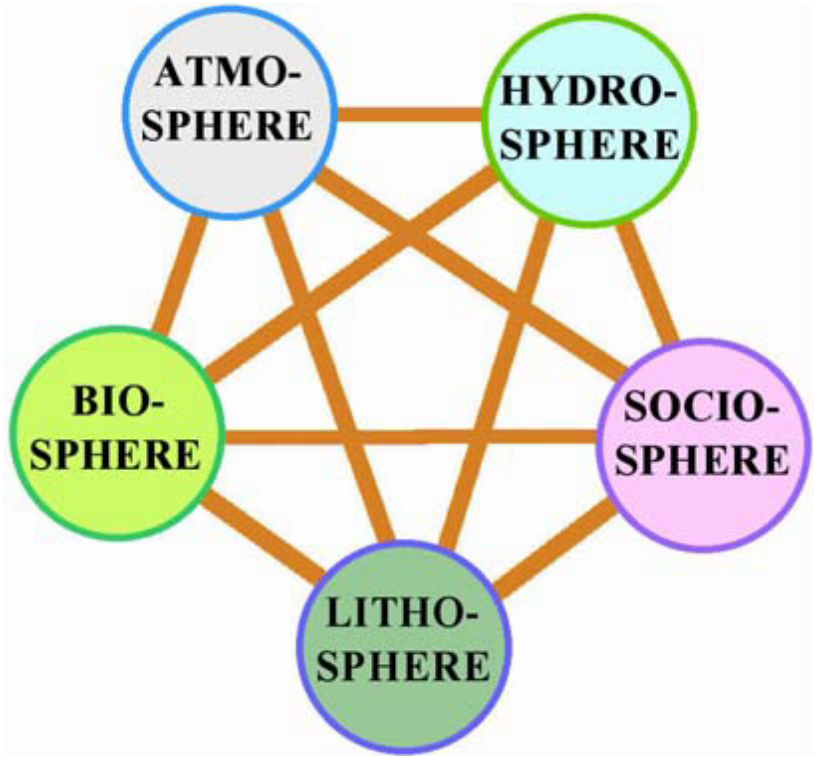

---------------------------
The Earth as five spheres in mutual interaction
[http://exergy.se]
---------------------------

This diagram may be taken to be a brief description of James Lovelock's Gaia concept extended to our times. The sociosphere affects human decisions on use of Earth's resources which may alter irretrievably the other spheres. Geologists have proposed that this effect is now large enough to be formalized in naming the present geologic epoch Anthropocene. Whether the biosphere of the far distant future will have place for humanity is a moot question. The same concept is illustrated below somewhat differently.

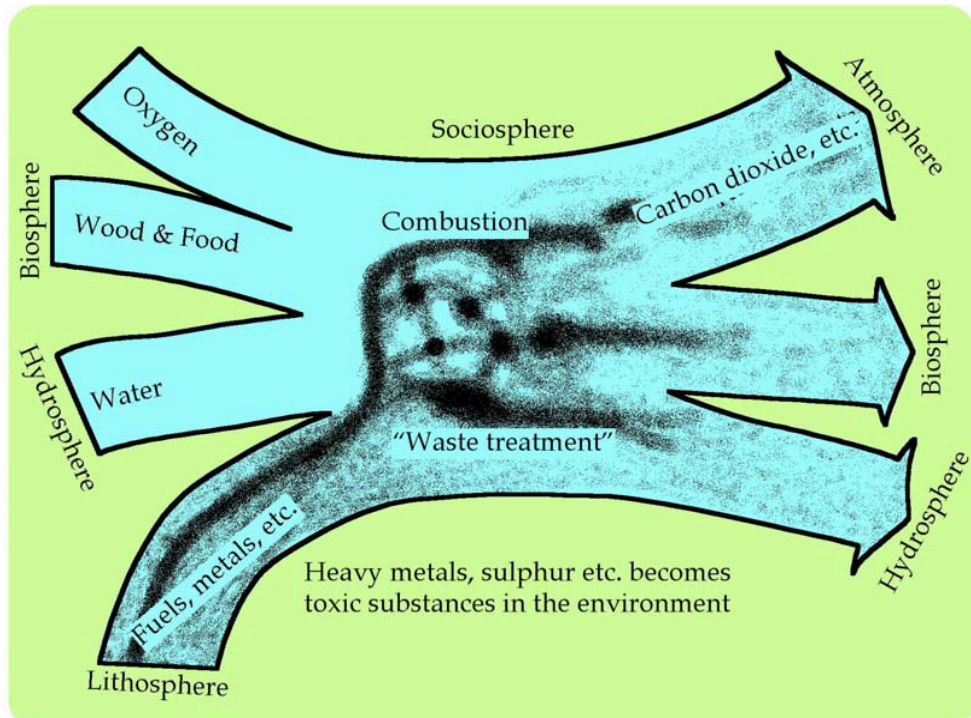

**Resource depletion and environmental destruction are two sides of the same problem.**
(From Göran Wall preprint 2011 Exergy and Sustainable Development p. 20 – cf [http://exergy.se].)

Two diagrams below illustrate that traditional farming was sustainable, while modern industrial farming has irretrievably upset Earth's nitrogen cycle balance.

---

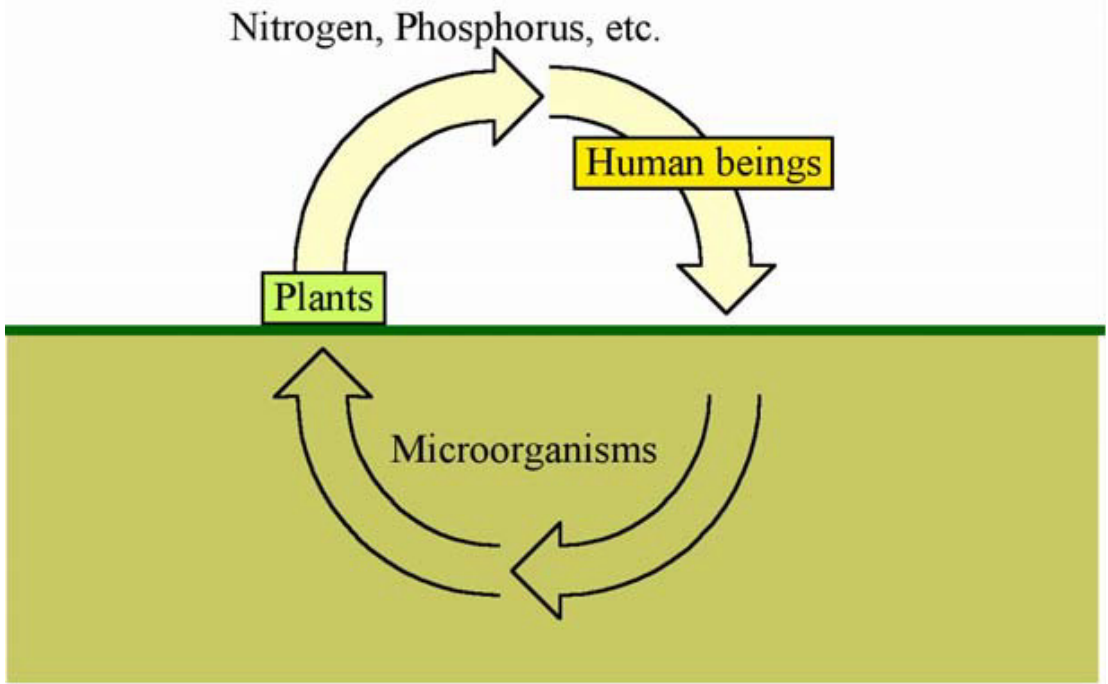

Traditional farming with recycling of matter in order to be sustainable.
**(From Göran Wall preprint 2011 Exergy and Sustainable Development p. 30 – cf [http://exergy.se].)**

---

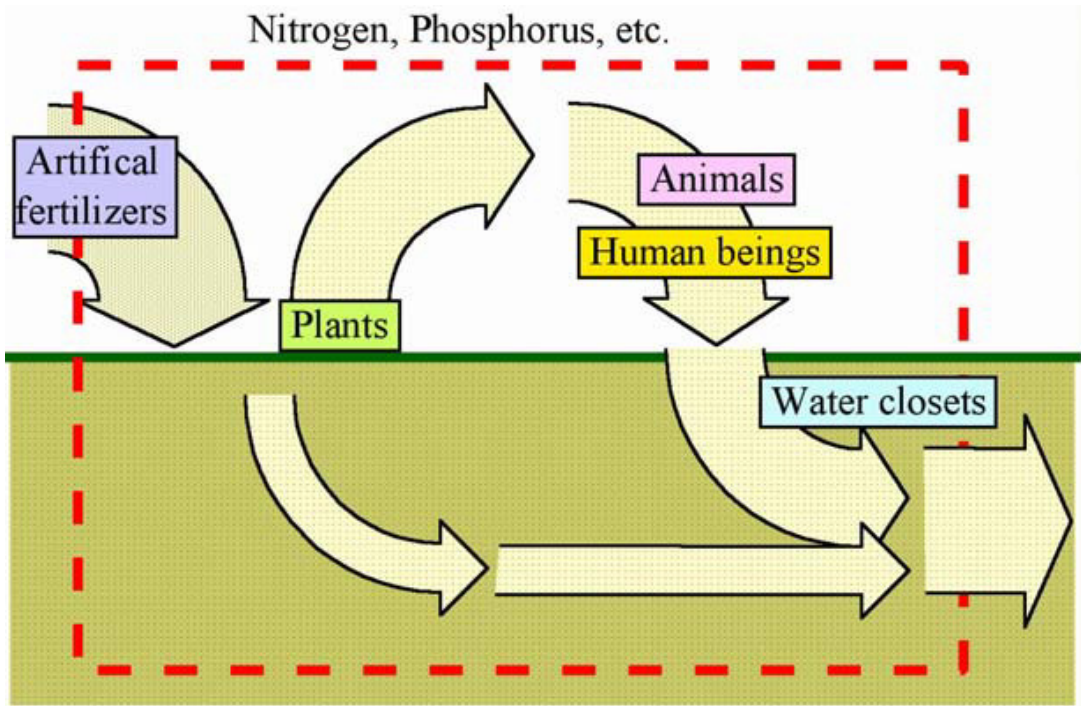

Modern industrial farming with artificial fertilizers and water closets.
**(From Göran Wall preprint 2011 Exergy and Sustainable Development p. 31 – cf [http://exergy.se].)**

---

Very clear dead zones exist near many coasts on Earth, especially as visible from satellites. Here the only biota are algae feeding on nitrogenous effluents coming into sea from land via surface rivers as well as underground streams. This toxicity has been a result of decades of using nitrogenous fertilizers for short-term gains in produce. Before this, agriculture was sustained by age-old experience of organic farming, and nitrogen cycle was in balance.

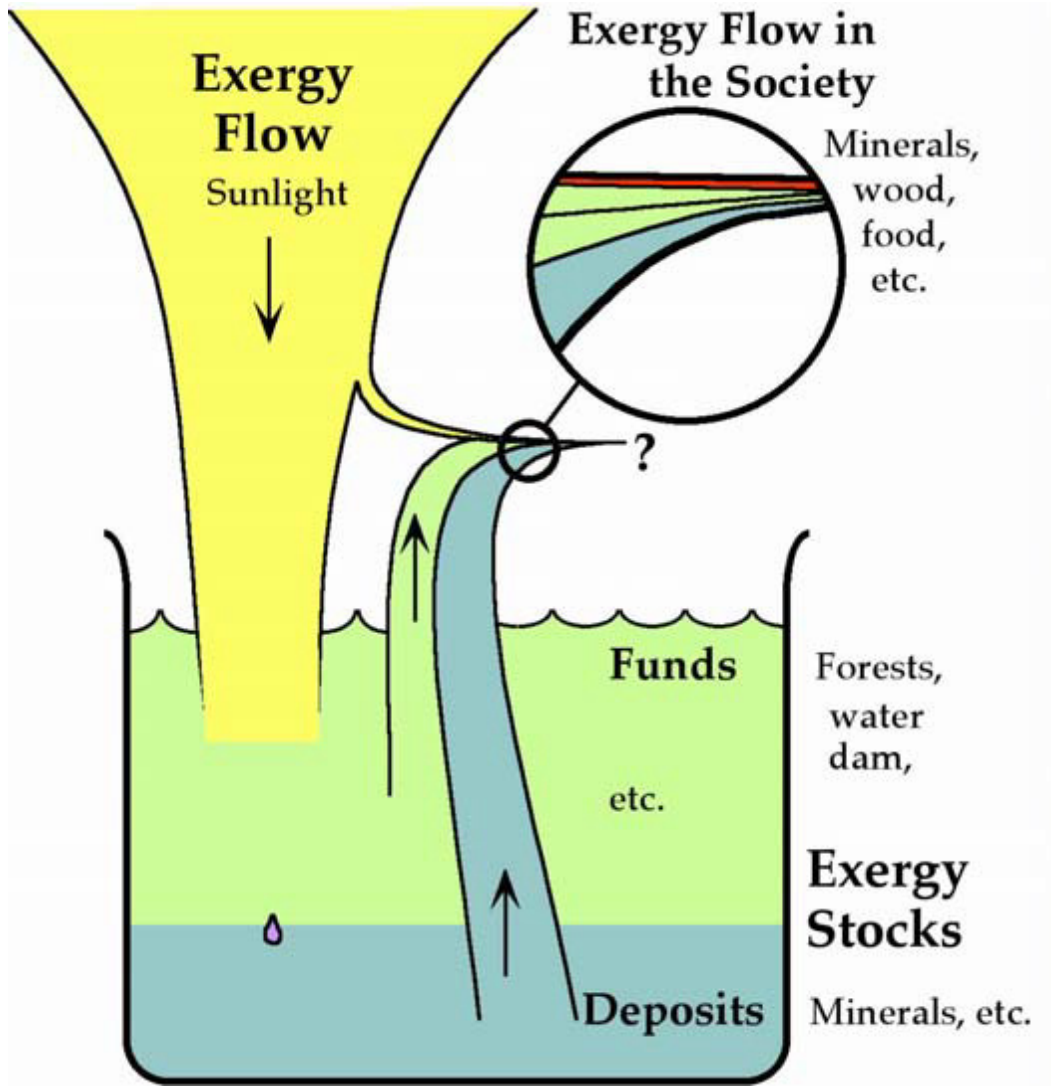

The exergy flow from the Sun, and the exergy stocks on Earth create the resource base for human societies on Earth
**(From Göran Wall preprint 2011 Exergy and Sustainable Development p. 21 – cf [http://exergy.se].)**

**APPLICABILITY OF EXERGY TO ECOLOGY & SUSTAINABILITY**

The application of exergy analysis for ecological questions and sustainability matters has just begun. Please see references [17]-[29] for examples. One important point made in ref. [27] is that instead of categorizing resources as renewable or non-renewable, it is more practical to consider renewability over daily, short-cycle (say, year), long-cycle (a decade) or even larger geological timescales (the largest possible on Earth).

**CONCLUSION**

In this article, after giving a brief history of thermodynamics and presenting its essential concepts, exergy is introduced, defined and elaborated a little. An example of its use in a thermal power plant is then displayed. As a pictorial survey of the exergy concept applied to natural processes on Earth, many figures and diagrams from [http://exergy.se] are then presented, along with commentary on each.

**Acknowledgements**

We are grateful to organizers of Bengaluru Humboldt Kolleg for allowing the second author (DGB) to present this paper. Finally, DGB recalls fondly Prof. B. B. Parulekar of IIT Bombay Mechanical Engineering Department for teaching Refrigeration and Air Conditioning very fruitfully using the exergy concept. In fact, it was this strong memory which reminded him of exergy as very appropriate for today's world, and he was pleasantly surprised to find that we are among many friends with this same consideration!